\documentclass[11pt]{article}
 \pdfoutput=1
 \usepackage[margin=1.25in]{geometry}
\usepackage{amsmath}
\usepackage{times}
\usepackage{graphicx}
\graphicspath{{./figures/png/}}
\usepackage{multirow}
\usepackage{libertine}
\usepackage[libertine]{newtxmath}
\usepackage[authoryear]{natbib}
\usepackage[hidelinks]{hyperref}
\usepackage{float}
\usepackage{setspace}
\usepackage{setspace}
\usepackage{rotating}
\usepackage{bbm}
\usepackage{latexsym}
\usepackage{tabularx}
\usepackage{rotating}

\usepackage[font=footnotesize,labelfont=bf]{caption}
\newcommand{\specialcell}[2][c]{%
  \begin{tabular}[#1]{@{}l@{}}#2\end{tabular}}

\usepackage{xcolor}
\definecolor{Gray}{gray}{.25}

\title{Scaling Spike Detection and Sorting for Next Generation Electrophysiology}
\author{Matthias H. Hennig${}^\text{a}$, Cole Hurwitz${}^\text{a}$ and Martino Sorbaro${}^\text{a,b}$}

\date{\today}

\begin{document}

\maketitle\vspace{-5mm}

{\footnotesize
\noindent {{\bf a} Institute for Adaptive Neural Computation,  School of Informatics, University of Edinburgh, Informatics Forum, 10 Crichton Street, Edinburgh EH8 9AB, Scotland, United Kingdom }
\noindent {\bf b} {School of Computer Science and Communication, KTH Royal Institute of Technology, Lindstedtsv\"agen 5, 114 28 Stockholm, Sweden}}

\paragraph*{Abstract}
Reliable spike detection and sorting, the process of assigning each detected spike to its originating neuron, is an essential step in the analysis of extracellular electrical recordings from neurons. The volume and complexity of the data from recently developed large scale, high density microelectrode arrays and probes, which allow recording from thousands of channels simultaneously, substantially complicate this task conceptually and computationally. This chapter provides a summary and discussion of recently developed methods to tackle these challenges, and discuss the important aspect of algorithm validation, and assessment of detection and sorting quality.

\section{Introduction}
\label{sec:intro}

Extracellular electrical recording of neural activity is an essential tool in neuroscience. If an electrode is placed sufficiently close to a spiking neuron, the extracellular potential recorded often contains a clear, readily detectable signature of the action potential. As extracellular electrodes do not interfere with neural function, such recordings provide an unbiased and precise record of the functioning of intact neural circuits.

Recent progress in CMOS technology (Complementary metal-oxide semiconductor technology for low power integrated circuits) has provided systems that allow recording from thousands of closely spaced channels simultaneously with ever increasing density and sampling rates. With this technology, it becomes possible to reliably monitor several thousand neurons simultaneously both {\em in vitro} and {\em in vivo} \citep{Eversmann2003,Berdondini2005a,Frey2010,Ballini2014,Muller2015,yuan2016microelectrode,lopez201622,jun2017fully,dimitriadis2018not}. This is a significant advancement as it enables, for the first time, the systematic investigation of interactions between neurons in large circuits. Understanding these interactions will contribute to learning more about how neural circuitry is altered by cellular changes in diseases, injury, and during pharmacological interventions.

To appreciate the advantages of recording the activity of many neurons, it is important to emphasize that neural circuits are usually highly diverse and heterogeneous \citep{Hromadka2008,buzsaki2014log,Panas2015}. Not only do they consist of different neuron types, but even within groups of neurons of the same type, the firing rates may differ by orders of magnitude. This observation has been made consistently {\em in vitro} and {\em in vivo}, and it stands to reason that this has biological relevance. Conventional technologies, which allow simultaneous recording of a handful (rarely more than a hundred) of neurons, severely under-sample highly heterogeneous populations. If the recorded neurons are not representative of the whole population,  both experimental accuracy and reproducibility between experiments will be negatively affected. Moreover, dense recording systems increase the fraction of neurons isolated in a local population, to a level that was, so far, only accessible with calcium imaging.


A further advantage of recording many neurons at once is that it can be an effective way of probing neural excitability and connectivity, using functional interactions as a proxy measure for the effects of synaptic interactions. {\em In vitro} assays are particularly suited for investigation of functional interactions, as they can be augmented with stimulation, fluorescent labeling and targeted optogenetic stimulation \citep{zhang2009integrated,Obien2015}. A combination of dense multielectrode arrays and imaging technologies could allow phenotyping at the level of single cells, potentially in combination with further modalities such as gene expression profiling. The high yield of such approaches thus provides entirely new possibilities for systematic assessment of the roles of different genotypes and of drug effects.

The analysis of single neuron activity requires the correct assignment of each detected spike to the originating neuron, a process called {\em spike sorting}.
In this chapter, we will provide an overview of the most frequently employed methods for the spike sorting for large-scale, dense multielectrode arrays. While many of the issues discussed will also apply to dense {\em in vivo} probes, the focus is on {\em in vitro} arrays, because they typically provide a large surface area evenly covered with recording channels, which is advantageous for spike sorting. A major additional challenge in {\em in vivo} recordings is tissue movement, which causes the signals of neurons to drift over time. For an excellent review of the challenges encountered {\em in vivo}, and of methodology for conventional recording devices with fewer channels, the reader may consult \citet{Rey2015}.

In the first section, we will discuss in more detail the technical and practical issues that are introduced when moving from conventional devices with tens of channels to larger, more dense systems. Next, we will introduce the main components of modern spike sorting pipelines, and then discuss each component and existing algorithms in detail. Finally, we will provide an overview of approaches for validation of the quality of these algorithms.

\section{Challenges for large-scale spike sorting}
\label{sec:chall}

On both conventional and high density recording devices, electrodes will usually pick up the activity of multiple neurons. While it is possible to directly analyse the multi-unit activity (MUA) from each channel, spike sorting is required to resolve single-unit activity (SUA). Spike sorting resembles the classic ``cocktail party'' problem: to isolate the voice of a single speaker in a crowd of people. Since the recorded spike waveforms differ in shape and amplitude among neurons, the resulting signal can be de-mixed using either dimensionality reduction paired with clustering or spike templates along with template matching. These approaches have been successfully employed on conventional devices with few, spatially well separated channels. On large-scale, dense arrays, however, these traditional methods become more difficult both computationally and algorithmically. Instead of finding a single voice in a crowd, the challenge is to isolate the voices of thousands of speakers in a room equipped with thousands of microphones. Overcoming this challenge is imperative as wrong assignments can severely bias subsequent analysis of neuronal populations \citep{Ventura2012}.

Spike sorting is a tractable problem for conventional extracellular recordings as it is commonly done for each recording channel separately. In this case, only a small number of neurons are expected to contribute to the signal on each channel, which allows the use of precise, but computationally more costly algorithms. Also, most existing algorithms for spike sorting still include an element of manual intervention to adjust or improve sorting results. These traditional algorithms struggle when faced with large-scale, dense arrays.

On dense arrays, a single action-potential from a neuron is visible on multiple, nearby channels. As a result, spike sorting on single channels is no longer appropriate. Removing duplicate events is feasible in principle, but becomes challenging when nearby neurons are firing with high synchrony. Poor treatment of duplicate removal can lead to false exclusions of action potentials or retention of multiple spikes from the same action potential.

Conventional spike sorting algorithms also struggle with the sheer volume of data large-scale arrays produce. For instance, a recording from 4,096 channels with 18kHz sampling rate yields about 140 megabytes per second, or over 8 gigabytes per minute. Simply reading this data volume from hard disk into memory for analysis can be a severe bottleneck in any spike sorting pipeline. In addition, the massive data volume prevents extensive manual curation of spike sorting results. Highly automated pipelines with minimal need for intervention are needed to overcome these challenges and to fully exploit the capabilities of dense arrays.


\section{From raw data to single neuron activity}

\begin{figure}
\centering
\includegraphics[scale=.4]{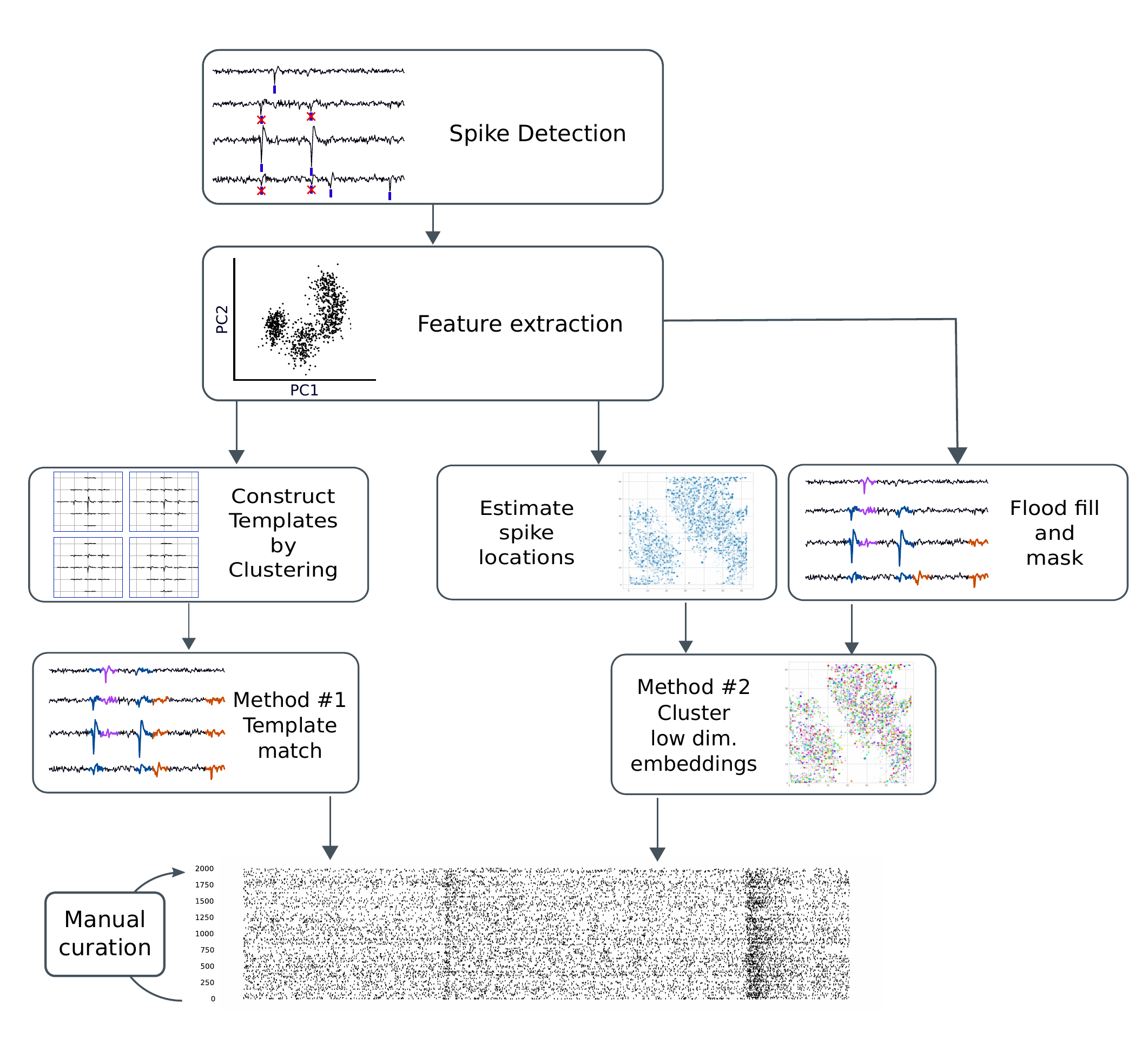}
\caption{Schematic overview of existing spike sorting pipelines for high density microelectrode arrays. Following detection, either neuron templates are formed based on the spatio-temporal event footprints and then used to detect these units in a second pass, or current sources are estimated and clustered together with waveform features, or a mask is created to restrict clustering to channels with a detectable signal. The output consists of a list of spike time stamps for each identified neuron, which often has to be corrected in a final manual curation step.}
\label{fig:1}
\end{figure}

A typical spike sorting pipeline begins with the detection of candidate events followed by some method of assigning these events to specific neurons. \citep{Lewicki1998,Rey2015}. On large-scale arrays, two approaches have emerged as particularly suitable. One method is based on creating spike templates and then performing template matching. The other method relies on feature extraction and clustering, using both the spike shape and estimated location of the event. A summary of the steps required to obtain sorted spikes from raw data is shown in Figure \ref{fig:1}. Each of these steps is discussed in more detail below.

\subsection{Spike detection}
\label{sec:detect}

Spikes in the raw signal take the form of biphasic deflections from a baseline level. They can be found through detection of threshold crossings and by using additional shape parameters such as the presence of a biphasic shape as acceptance criteria. As the noise levels may vary among channels and over time, the threshold is usually defined relative to the noise level, which is estimated from portions of the raw signal that do not contain spikes. It is worth noting that signal fluctuations in extracellular data are typically highly non-Gaussian. As a result, a noise estimate based on percentiles is more accurate and also easier to obtain, as opposed to computing the signal variance \citep{fee1996variability,Muthmann2015}.



\begin{figure}
\centering
\includegraphics[scale=.65]{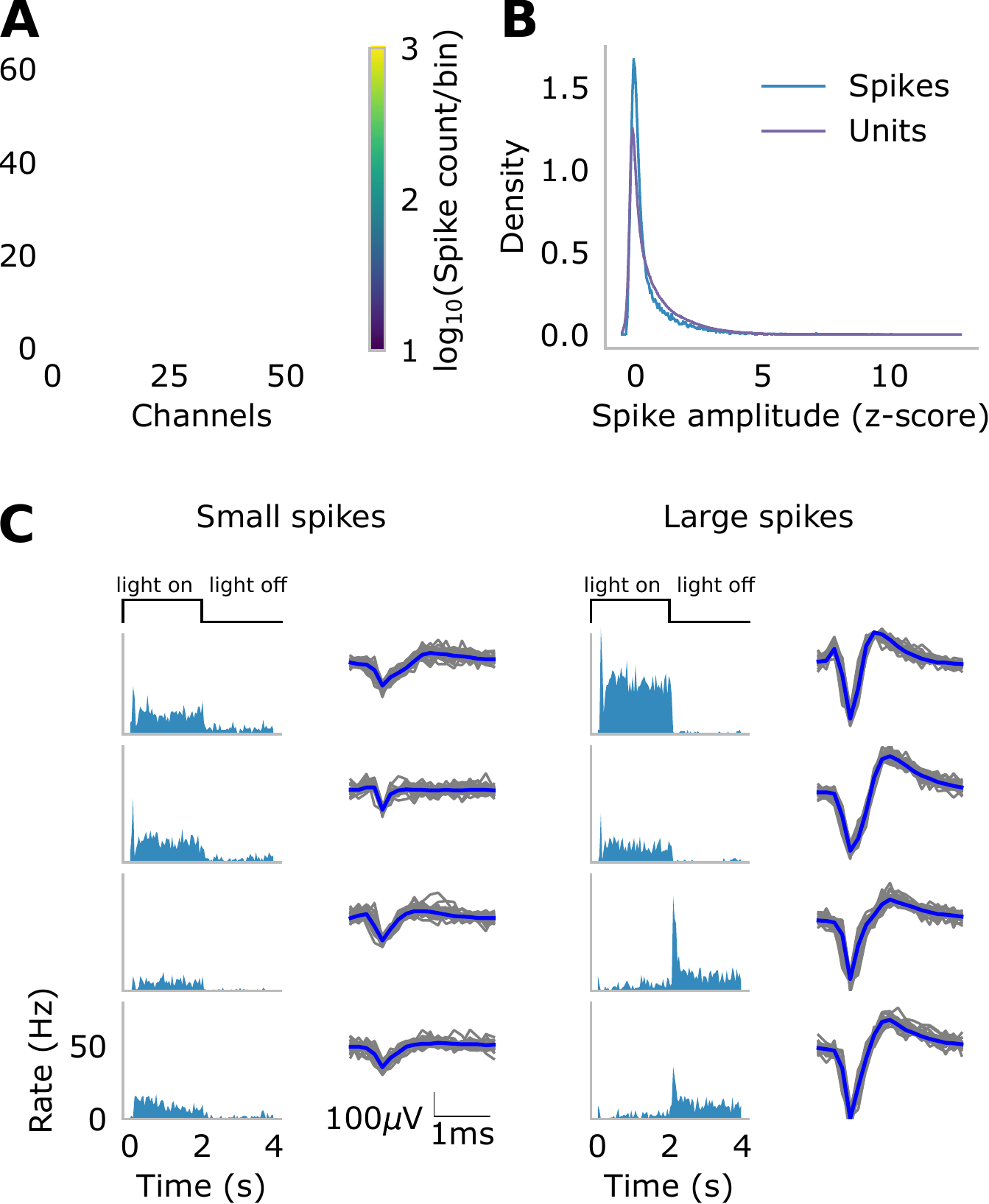}
\caption{Even small events detected on a high density array contain signatures of neural activity. A, Density plot of spatially binned spike counts, estimated from detected and spatially localised spikes using the method described by \citet{hilgen2017unsupervised}. Spikes were recorded from a light-stimulated mouse retina. Spike detection was performed with a low threshold, hence false positives were registered in areas where no neural activity was recorded, such as the optic disk on the centre left. B, No clear separation between spikes and noise is seen for recorded amplitudes, or for average amplitudes of units following spike sorting. C, Individual, randomly selected units with small (left) and large amplitudes (right) both show signatures of light stimulation during presentation of full field flashes. For each unit, the light-evoked peri-stimulus time histogram (left), and examples and the average of spike waveforms (right) are shown. The recording was contributed by Gerrit Hilgen and Evelyne Sernagor, University of Newcastle.}
\label{fig:2}
\end{figure}

The choice of the detection threshold determines which events are retained for further analysis. Spikes from well-detected neurons are easily identifiable, but deflections with amplitudes closer to the background noise level are harder to isolate. Since there is typically no clear-cut separation between spikes and noise, events detected close to the threshold may originate from neurons for which only an incomplete activity record can be obtained. The magnitude of electrical noise, which can be estimated when recording from an empty array, is usually much smaller than the magnitude of the fluctuations recorded in the absence of clearly visible spiking activity \citep{Muthmann2015}. This indicates that a large component of the recorded signal fluctuations are due to neural activity, such as neurons located further away from the electrode, or smaller events such as currents during synaptic transmission. The analysis of recordings from the retina shows that even very small detected signals may reflect activity that is typical of stimulus-evoked responses from retinal ganglion cells, hence carries signatures of neural activity rather than noise (Figure \ref{fig:2}).

As a result, the detection step significantly affects the subsequent isolation of single neuron activity. Choosing a high detection threshold is not an ideal solution as this will potentially leave valid spikes undetected. In contrast, a low threshold guarantees reliable detection of neurons with stronger signals, but also increases the fraction of false positives. As a good compromise, a strategy can be adopted to detect events with a low threshold, and to subsequently discard unreliable units. This can either be done after detection, for instance by using a classifier trained on true spikes and noise events obtained from channels not reporting neural activity \citep{hilgen2017unsupervised}, or by removing sorted units with a small number of spikes or poor clustering metric scores post spike sorting \citep{Hill2011}.

Recently, a new method for spike detection using a pre-trained neural network was introduced by \citet{lee2017yass}. This method was shown to outperform conventional threshold-based methods on simulated ground truth data, in particular by achieving a lower false positive rate. When run on a modern GPU, a neural-network based method is also much faster. This is a very promising avenue, although the considerations regarding detection thresholds outlined above still remain relevant.

\subsection{Dealing with duplicate spikes}
\label{sec:dup}

Unlike in conventional recordings, on dense arrays, spikes are detectable on multiple channels. These duplicate spikes pose two significant problems for traditional spike sorting algorithms. First, the amount of computation and memory used for processing each detected event increases with the number of duplicates. Second, the rate of misclassification in spike sorting potentially increases since each duplicate spike must be sorted into the same event.

To avoid the pitfalls associated with duplicate spikes, it is suggested to identify and remove duplicates during detection. One naive method for duplicate removal is to remove all but the largest amplitude spike in a radius that encompasses the spatial footprint of the event. This method will remove almost every duplicate event, but as the radius of duplicate removal increases, so does the number of spikes removed that are not associated with the original event. A more rigorous method for duplicate removal involves keeping the largest amplitude spikes and removing all spikes in a radius that have \textit{decayed in amplitude}. This method allows for the separating of near-synchronous events that are in the same spatial area of the array. Its success, however, relies on the assumption that the timing of spikes from the same event on nearby electrodes is almost identical and only weakly influenced by noise and that the signal spatially decays away from its current source \citep{Hagen2015}.

\subsection{Feature Extraction}
\label{sec:feat}

The relevant signal a spike causes in extracellular recordings lasts around 3\,ms, which, depending on the acquisition rate, may correspond to up to 90 data points per event. However, spike shapes are highly redundant and can be efficiently represented in a low dimensional space. Thus, an appropriate projection method can be used to compute a small number of features for each event, which can be more efficiently clustered than raw waveforms.

The most common feature extraction method for extracellular spikes is Principal Component Analysis (PCA), performed on whitened and peak-aligned spike waveforms. PCA finds principal components, or orthogonal basis vectors, whose directions maximize the variance in the data. Extracellular spikes can be summarized well by just 3-4 principal components, a manageable dimensionality for most clustering algorithms. \citep{adamos2008performance}. Other less frequently used methods include independent component analysis (ICA) \citep{hermle2004employing} and wavelet decomposition \citep{Quiroga2004}. A comparison of these methods showed that the performance of sorting algorithms depends not only on the feature extraction method employed, but also on the clustering algorithm \citep{Quiroga2004}. In practice, the comparably low computational cost and relative effectiveness of PCA in discriminating between different neurons and neuron types makes it particularly suitable for large scale recordings \citep{adamos2008performance}. To reduce memory load, the PCA decomposition can be evaluated for a subset of events from a large recording and all events can be projected along the chosen dimensions efficiently in batches \citep{hilgen2017unsupervised}.

\subsection{Clustering spatio-temporal event footprints}\label{sec:clustering}

There are five fundamental problems with the clustering phase. The first problem is that the extracellular waveform of neurons are known to change amplitude and shape during bursting \citep{fee1996variability}. The second problem is that some recorded waveforms are distorted by overlapping action potentials from synchronous, spatially-local events. This occurs frequently in dense arrays and usually exist as outliers in the chosen feature space. The third problem is that electrodes can drift in the extracellular medium, changing the relative position of each neuron to the electrodes. Drift distorts waveform shapes over the duration of the recording. The fourth problem is that the duplication of spikes over neighboring channels can lead to refractory period violations or misclassifications. The fifth and final problem is that the number of observed neurons is unknown, which requires the use of non-parametric clustering algorithms or requires the user to estimate the number of neurons for a parametric clustering algorithm.

The choice of the clustering algorithm will be determined by the speed and scalability considerations, by hypotheses over the typical shape of a cluster in this space, and by how well the algorithm can deal with the previously listed problems. Many spike sorting methods cluster by fitting Gaussian Mixture Models (GMMs), modelling the feature density profiles as a sum of Gaussians \citep{Harris2000,Rossant2015}, or by fitting a mixture of t-distributions \citep{Shoham2003111}. The unknown number of actual neurons can be introduced as a latent variable, and the inference problem be solved with the expectation-maximisation (EM) algorithm. Bayesian approaches, which also quantify parameter uncertainty, have also been introduced \citep{wood2008nonparametric}. These approaches, however, only perform well for single channels and are conceptually and computationally hard to scale up to large, full-chip datasets.

More recent clustering algorithms for spike sorting are density-based. Density-based algorithms generally detect peaks or high-density regions in the feature space that are separated by low-density regions. These algorithms are non-parametric, allowing the classification pipeline to be fully automatic, however, the number of clusters found can depend heavily on both hyper-parameters and the chosen feature space. Density-based clustering algorithms have been implemented for spike sorting with promising results \citep{hilgen2017unsupervised,chung2017fully}.

For dense arrays, an added complication arises since the information contained in event footprints cannot be used directly for sorting spikes, since it is unknown which channels contain responses of a single neuron and how many neurons cause the observed responses. The resulting combinatorial explosion can be dealt with in three ways:

\paragraph*{Masked clustering}
A straight-forward way to reduce the dimensionality of the clustering problem is to include only channels with detectable responses for each event. Classical expectation maximisation on a mixture model is then possible when the irrelevant parts of the data are masked out and replaced with a tractable noise model \citep{kadir2014high}. This strategy produces excellent results with the help of a semi-automated refinement step. \citep{Rossant2015}. A main limitation is, however, a super-linear scaling with the number of recordings channels, which makes it less suitable for the latest generations of large-scale arrays.

\paragraph*{Template matching}
Since the raw recorded signal can be linearly decomposed into a mixture of footprints from different neurons \citep{segev2004recording}, template matching has been a successful strategy for spike sorting, and implementations are available that scale up to thousands of channels \citep{pachitariu2016fast,lee2017yass,yger2018spike}. This approach has two steps. First a collection of spatio-temporal footprints is obtained in a single pass over the data and dimensionality reduction and clustering is used to build templates for single neurons. In a second pass, all events are assigned to the most likely template or combination of templates in the case of temporally overlapping events.

A major advantage of template matching that temporally overlapping spikes are naturally accounted for through addition of two relevant templates. This makes it very suitable for recordings with high firing rates and correlations between nearby neurons. A potential limitation is that neurons spiking at very low rates may remain undiscovered as no reliable template can be built through averaging. Moreover, current implementations require a final manual curation step. This is, however, simplified by correcting the assignment based on templates, which can be merged or split, rather than based on single events.

\paragraph{Spike localisation}
As explained above, the spatial spike footprint allows event localisation through an estimation of the barycentre from the peak event amplitudes in nearby channels. This produces density maps with clear, isolated peaks in event density, which represent spikes from single or multiple, nearby neurons (see Figure \ref{fig:2} for an example). A two-dimensional density map can be clustered very efficiently, and the combination of locations and waveform features obtained through dimensionality reduction allows successful separation of nearby neurons. Density-based clustering algorithms have been successfully employed to solve this task: \textsc{DPClus}, based on the identification of density peaks \citep{jun2017real}, \textsc{ISO-SPLIT} to grow uni-modal clusters from small seeds \citep{chung2017fully} and Mean Shift, which herds data points towards high-density areas \citep{hilgen2017unsupervised}.

Of all methods discussed here, spike localisation and clustering potentially has the best computational performance, since the actual computation is performed on a data set with much lower dimensionality than the original data \citep{hilgen2017unsupervised,jun2017real}. Because the number of dimensions in the clustering step has to be kept small, it also discards useful information. However, usually locations and spatio-temporal waveform features exhibit substantial redundancy \citep{hilgen2017unsupervised}, making this approach the most suitable for very large arrays.

\section{Evaluation}

The evaluation of spike detection and sorting quality is complicated by data volume and complexity, which makes both manual and automated curation challenging. It is however possible to assess the quality of an algorithm using data with ground truth annotation. Moreover, methods for post-hoc quality assessment of desirable properties of single units can be used to accept or reject units found through spike sorting.

Specifically, the desired result of a spike sorting pipeline to minimise the false detection of noise as spikes (false positives in detection), and the number of real spikes left undetected (false negatives in detection). Moreover, it should not assign spikes to the wrong neuron, hence it should minimise false positives and negatives in a cluster assignment.

When ground truth annotations are available, false positives and false negatives can be easily counted. A direct, but technically challenging method to obtain ground truth information, is the simultaneous recording of a single neuron, together with an array recording, which will then be analysed using the spike sorting algorithm in question. Three such data sets recorded with dense arrays are currently available, two from the rat cortex recorded {\em in vivo} \citep{neto2016validating,MarquesSmith2018}, and one from the mouse retina recorded {\em in vitro} \citep{yger2018spike}. In both cases, a single juxtacellular electrode placed very closely to the array reliably recorded all spikes from a single neuron. A systematic analysis of spike sorting has shown a clear relationship between measured spike amplitude and classification accuracy with errors strongly increasing for events smaller than 50 mV \citep{yger2018spike}. This important result can help motivate exclusion of units with weaker signals.

Ground truth for spike sorting can also be produced by simulations. Recently, it has become possible to simulate a complete biophysical forward model for recorded extracellular potentials in neural tissue \citep{Hagen2015}. This has produced several data sets that are now used to benchmark spike sorting algorithms (see e.g. \cite{lee2017yass}). In another study, ground truth data was generated by superimposing synthetic spikes onto a recording from an empty array. This data was used to evaluate the effect of noise on event localisation accuracy and to discover that localisation is inevitably a trade-off between position uncertainty and bias \citep{Muthmann2015}. It is an open question, however, how well results collected from simulated data generalise, since the precise noise model, which may differ between recording systems, impacts spike sorting algorithm performance \citep{Muthmann2015}.

Finally, for cases where no ground truth data is available, \citet{Hill2011} proposed a set of metrics that should accompany all spike sorting methods as an evaluation of their reliability. Their metrics, applied \emph{a posteriori}, are based on different features of the sorted dataset, which can be summarised as follows:
\begin{itemize}
	\item The \textbf{waveforms} in each cluster. The average waveform can present non-biological features, hinting that the cluster may be a collection of wrongly detected events. Additionally, if properties of the typical waveform change over time, this may be a sign of neurons drifting away from their initial position on the chip. Finally, anomalous variability of each feature above the noise level may be a sign that multiple neurons contributed to the same cluster.
  \item The \textbf{times} of all spikes in a cluster. Violations of the refractory period show the cluster contains false positives: these can be studied via the autocorrelation function or inter-spike histogram of each cluster.
  \item The \textbf{amplitudes} of action potentials. A sharp drop in the amplitude distribution, caused by the detection threshold, signifies that the latter has introduced an artificial bias.
  \item The \textbf{separation} between pairs of clusters. Ample, sharp interfaces between clusters mean the properties of each neuron's spikes overlap in the selected feature space. If this occurs, there will be a theoretical minimum of false positives and negatives due to the incorrect assignment of events to the wrong cluster.
\end{itemize}
The last point can be evaluated by re-examining a group of clustered neurons with a mixture model (usually Gaussian), which can be fit using more features than the original algorithm. Assuming that this fit is at least as reliable as the original sorting algorithm, a comparison of the two assignments is informative regarding the reliability of each unit. A statistic summarizing all these tests can then be used to exclude events and units {\em post hoc}. Using this method, detection and clustering parameters do not have to be adjusted carefully prior to each analysis.

\section{Outlook}
\label{subsec:conc}

In this chapter, we discussed the existing methodology for recovering single neuron activity from high density recordings and the challenges and problems that each approach faces. Six freely available spike sorting pipelines for large-scale extracellular arrays and the methods they use are summarised in Table \ref{tab:sortingmethods}. For more information on their unique advantages and disadvantages, please review their associated references.


\begin{table}
\begin{tabularx}{\textwidth}{l|c|X}
  	\textbf{Name and reference} & \textbf{Method} & \textbf{Notes} \\
    \hline
    \hline

    \specialcell[c]{Kilosort (\cite{pachitariu2016fast}) \\ github.com/cortex-lab/KiloSort}& TM &  GPU support; MATLAB based; semi-automated final curation. \\
    \hline
    \specialcell[c]{YASS (\cite{lee2017yass}) \\ yass.readthedocs.io} & TM &   Neural network-based detection (GPU); outlier triaging; template matching;  clustering.\\
    \hline
    \specialcell[c]{Herding Spikes (\cite{hilgen2017unsupervised}) \\ github.com/mhhennig/HS2} & SL+D & Fast and scalable; tested on multiple array geometries \\
    \hline
    \specialcell[c]{MountainSort (\cite{chung2017fully}) \\ github.com/flatironinstitute/mountainsort}  & D & Fully automatic; scalable; graphical user interface; unique clustering method \\
    \hline
    \specialcell[c]{JRCLUST (\cite{jun2017real}) \\ jrclust.org} & SL+D & Probe drift correction; GPU support. \\
    \hline
    \specialcell[c]{SpyKING CIRCUS (\cite{yger2018spike}) \\ spyking-circus.rtfd.org}  & TM & GPU support; tested on many datasets; robust to overlapping spikes; graphical user interface.  \\
    \hline
\end{tabularx}

\caption{Summary of the most recent spike sorting methods developed for large, dense arrays. For a summary of older algorithms --mostly for smaller, sparser arrays -- see \cite{bestel2012novel}. TM = Template Matching; SL = Spike Localisation; D = Density-based clustering (see section \ref{sec:clustering})}
\label{tab:sortingmethods}
\end{table}

Since inaccurate detection and sorting can influence subsequent analysis of neural populations  \citep{Ventura2012}, manual curation steps are often still required to guarantee good data quality. However, the recent methods we summarised in this chapter take significant steps in increasing the speed, automatisation, and accuracy of the spike sorting pipeline. Looking forward, it may be possible to apply novel machine learning techniques to improving spike sorting. This has already been put into practice with a recent spike sorting algorithm where a neural network is used to improve detection of neural events \citep{lee2017yass}. Although neural networks are showing promising results in detection, it may be possible to find new breakthroughs in both feature extraction and in classification using these methods. Moreover, a neural network approach may have the potential of encompassing all of the spike sorting steps within a single model. A challenge when using these machine learning algorithms, however, is the difficulty of obtaining ground truth data, which is poorly available and usually under specific experimental conditions that may not generalize to other data sets.

Increased automation also means more work is needed in developing reliable methods for validation and quality control of spike sorting results. The introduction of synthetic \citep{Hagen2015} and experimetal ground truth datasets \citep{neto2016validating,yger2018spike} is an important step forward in this direction. A standardisation, both of the sorting pipeline and of its evaluation, should be considered among the next objectives of the spike sorting community. A joint effort should be made in order to guarantee that methods are intuitive to use and results are easy to compare.

\end{document}